# Screw-Dislocation-Driven Growth of Two-Dimensional Few-Layer and Pyramid-Like WSe$_2$ by Sulfur-Assisted Chemical Vapor Deposition

Liang Chen[+], Bilu Liu[+], Ahmad N. Abbas, Yuqiang Ma, Xin Fang, Yihang Liu, Chongwu Zhou*

Department of Electrical Engineering, University of Southern California, Los Angeles, California, 90089, United States

[+] Equal contribution.

E-mail: chongwuz@usc.edu

**Abstract.** Two-dimensional (2D) layered tungsten diselenides (WSe$_2$) material has recently drawn a lot of attention due to its unique optoelectronic properties and ambipolar transport behavior. However, direct chemical vapor deposition (CVD) synthesis of 2D WSe$_2$ is not as straightforward as other 2D materials due to the low reactivity between reactants in WSe$_2$ synthesis. In addition, the growth mechanism of WSe$_2$ in such CVD process remains unclear. Here we report the observation of a screw-dislocation-driven (SDD) spiral growth of 2D WSe$_2$ flakes and pyramid-like structures using a sulfur-assisted CVD method. Few-layer and pyramid-like WSe$_2$ flakes instead of monolayer were synthesized by introducing a small amount of sulfur as a reducer to help the selenization of WO$_3$, which is the precursor of tungsten. Clear observations of steps, helical fringes, and herring-bone contours under atomic force microscope characterization reveal the existence of screw dislocations in the as-grown WSe$_2$. The generation and propagation mechanisms of screw dislocations during the growth of $WSe_2$ were discussed. Back-gated field-effect transistors were made on these 2D $WSe_2$ materials, which show on/off current ratios of $10^6$ and mobility up to 44 $cm^2$/V·s.

Recently, two-dimensional (2D) layered materials beyond graphene have attracted huge amounts of attention especially for those transition metal dichalcogenides (TMDCs) with formula of $MX_2$ (M = Mo, W; X = S, Se).[1-4] Such TMDCs are also layered materials like graphene coupled by weak van der Waals forces between adjacent layers. But more than that, each TMDC layer contains three covalent bonded X-M-X atomic layers. Besides the unique structure, these materials exhibit interesting properties when the thickness goes down to monolayer or few-layers,[2, 5-7] making them as good candidates for advanced electronics,[8-11] optoelectronic devices,[12-15] energy storage devices,[16] and electrocatalysts.[17] So far, a lot of research efforts have been devoted to these materials focusing on material synthesis,[2, 18-23] characterization,[24, 25] fundamental property studies,[4, 26-28] and applications.[11-15, 29, 30] Among those four materials ($MoS_2$, $MoSe_2$, $WS_2$, and $WSe_2$), $MoS_2$ is the most heavily studied one. On the other hand, $WSe_2$ is not well studied until some recent papers.[13-15] Compared to $MoS_2$, $WSe_2$ possesses a smaller bandgap and it exhibits ambipolar transport phenomenon.[13-15]

Chemical vapor deposition (CVD) is a widely used method for TMDC synthesis.[18-21] Early in 2012, Balendhran and Kalantar-zadeh *et al.* have shown the preparation of thin $MoS_2$ flakes by evaporation of sulfur and $MoO_3$, following by exfoliation.[31] Although CVD growth of $MoS_2$ has been well developed, only a few papers reported the successful growth of 2D $WSe_2$ structures.[32-34] The difficulty of CVD growth of $WSe_2$ is believed to originate from the low reactivity of selenium.[20, 32] For example, Huang *et al.* have reported the synthesis of large area monolayer $WSe_2$ on sapphire under low-pressure CVD, and they found that adding $H_2$ as an additional reducing reactant is a must for successful $WSe_2$ synthesis.[32] Few-layer $WSe_2$ synthesis was also reported recently by Lin *et al.* using graphene as an epitaxial substrate.[34] Nevertheless, the growth mechanism of $WSe_2$ is still not clear. And more efforts should be devoted to further explore the controlled synthesis of $WSe_2$. Here we report a sulfur-assisted CVD method to grow few-layer and pyramid-like $WSe_2$ flakes following a screw-dislocation-driven (SDD) growth fashion. Such SDD growth is a universal growth mechanism in nanomaterials including 1D nanotubes,[35, 36] 1D nanowires,[37-39] and 2D nanoplates.[40, 41] However, it has not been reported in CVD synthesis of 2D layered materials. We believe this unique growth process in our method is due to low concentration of the reactants, and accordingly, relevant models are proposed to understand the SDD growth process in the CVD synthesis of $WSe_2$ flakes.

**Results and Discussions**

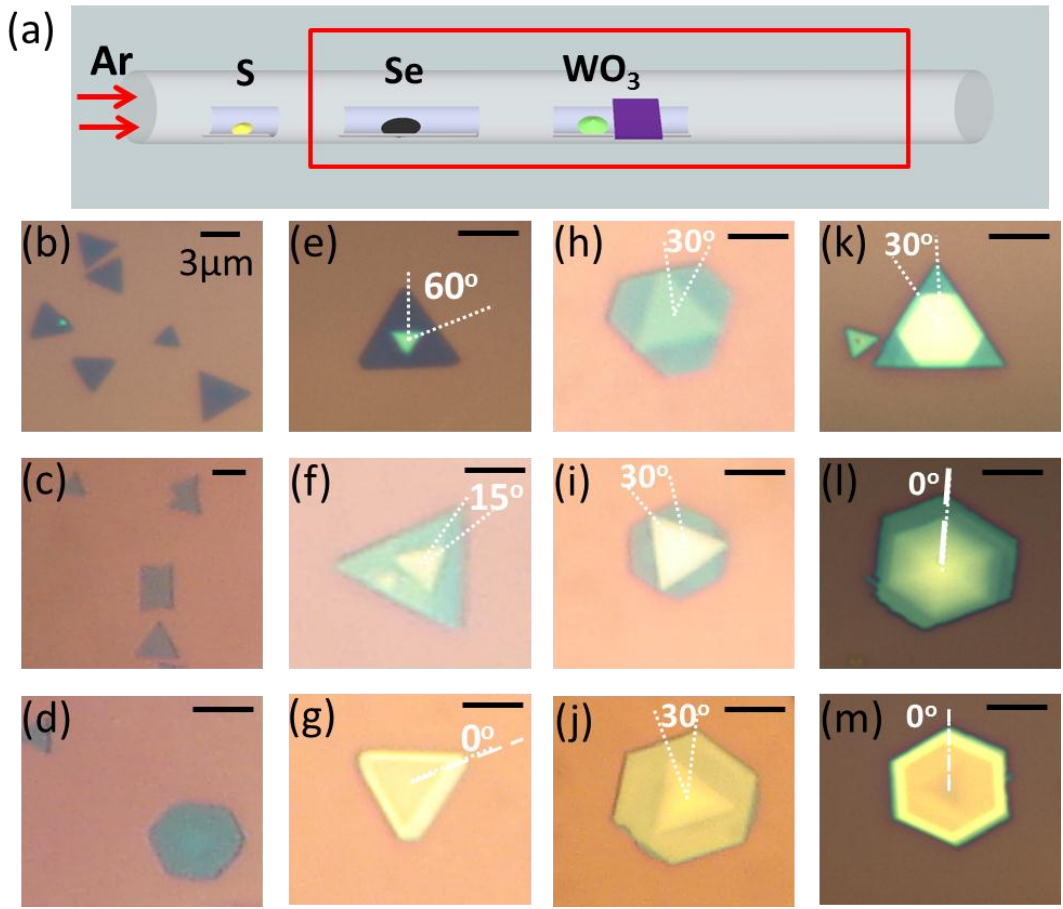

**Figure 1. CVD set up and optical microscopy characterization of CVD-grown few-layer and pyramid-like WSe$_2$ flakes. (a) A schematic diagram shows the CVD setup for the sulfur-assisted WSe$_2$ growth. (b, c, d) Optical microscopy images of thin WSe$_2$ flakes with different shapes. (e-m) Optical microscopy images of thick WSe$_2$ flakes with different stacking morphologies. The color contrasts reflect the thickness variations both from sample to sample and from center to edge within each sample. The white dotted lines indicate the stacking angles between top and bottom features in thick flakes. The scale bars are 3 μm for all images.**

Figure 1a illustrates the CVD setup we used for the synthesis of few-layer and pyramid-like WSe$_2$ flakes in this study. Similar to recent CVD methods for TMDC synthesis,[30,32] a quartz boat containing WO$_3$ powders with a Si/SiO$_2$ substrate sitting on top was placed in the middle of a 1-inch tube furnace. The temperature of WO$_3$ varied from 875 °C to 925 °C. Selenium and sulfur powders were loaded in two separated quartz boats and put at the upstream region. Note that only very small amounts of sulfur powders were used and they were intentionally put at a position with temperature below the melting point of sulfur to minimize their sublimation. After the growth, the SDD grown flakes can be found everywhere on the substrate. But the nucleation density is higher at positions closer to the WO$_3$ powders. More experimental details can be found in the Methods section.

By adjusting the growth temperature and time, a variety of WSe$_2$ flakes with different thicknesses and morphologies were synthesized (Figures 1b-1m), including few-layer triangles, few-layer hexagons, thick triangles, and thick hexagons. Figures 1b, 1c, and 1d are optical microscopy images of thin flakes synthesized at temperatures ranging from 875 °C to 900 °C. Besides the triangular and merged triangular flakes that are commonly observed in CVD-grown TMDCs,[25] hexagonal flakes are also occasionally found here. The lateral sizes of these flakes are ranging from 3 μm to 5 μm. When the growth temperature increased to 925 °C, we found that the flakes were predominantly thick ones based on optical microscopy observations. This is consistent with recent results showing that additional layers would grow at high temperature during CVD synthesis of MoS$_2$ and WSe$_2$.[32, 42] In our experiments, we found that the growth window of WSe$_2$ is narrow, and temperature plays the most important role in determining the thickness and shape of the materials. In general, both thin and thick flakes coexist at a temperature of 900 °C (Supporting Information Figure S1a), while at relative high temperatures (above 925 °C), all the flakes are thick ones with heights over 10 nm (Figure S1b). We did a statistical analysis of the samples grown at 900 °C based on Figure S1a, and the results show that 78% are thin flakes and 22% are thick flakes (>10 nm). Temperature also plays a crucial role in affecting the shapes of as-grown materials. Most flakes are triangular when the growth temperature is below 900 °C. After the growth temperature rises to 925 °C, both triangular flakes and hexagonal flakes exist. Statistical studies on the samples grown at 925 °C shows that 96% are triangular and 4% are hexagonal. We note that this ratio may vary from location to location on a substrate. Very interestingly, most thick WSe$_2$ flakes in our products have intriguing terrace-like morphologies with different stacking angles and shapes, as exhibited from Figures 1e to 1m. The color contrasts of these images in Figure 1 reflect the differences of thickness among the WSe$_2$ samples. Based on the optical microscope inspections, the thicknesses of these thick flakes decrease from centers to edges, indicating the formation of pyramid-like structures. To be noticed, although Figure 1 shows all possible morphologies observed during the experiments, some stacking types do appear more frequently than others. Detailed results will be discussed later.

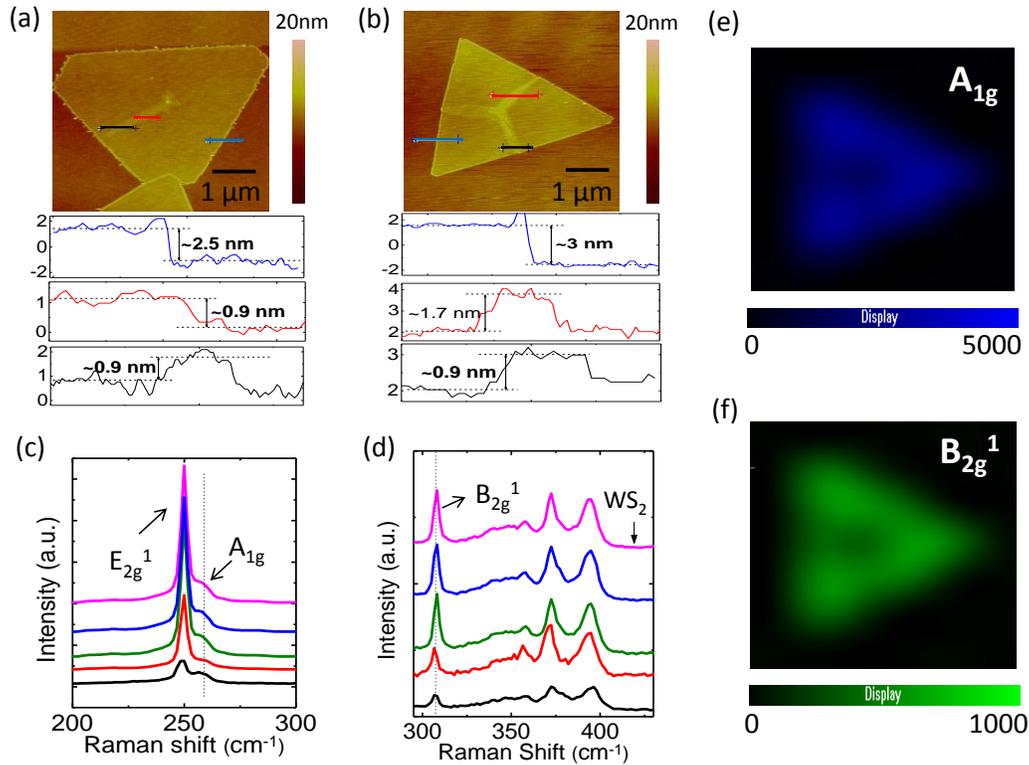

**Figure 2.** AFM and Raman characterization of thin $WSe_2$ flakes. (a, b) AFM images along with cross section height profiles of two thin $WSe_2$ flakes with ribbon-like features on top. The bottom layers are 2.5 nm and 3 nm in height, corresponding to around tri-layer $WSe_2$. The heights of ribbon layers are ~0.9 nm or ~1.7 nm, corresponding to one or two layers of $WSe_2$. (c, d) Raman spectra of several thin $WSe_2$ flakes. Characteristic peaks of $E_{2g}^1$, $A_{1g}$, and $B_{2g}^1$ modes of $WSe_2$ were detected while the $A_{1g}$ peak of $WS_2$ was not observed. (e, f) Raman intensity mapping of $A_{1g}$ (259 cm$^{-1}$) and $B_{2g}^1$ peaks (309 cm$^{-1}$) of the same $WSe_2$ flake shown in Figure 2(b). The excitation wavelength of laser is 532 nm during Raman measurements.

To further explore the detailed structures of thin and stacked thick $WSe_2$ flakes, we performed systematical atomic force microscopy (AFM) and Raman studies. AFM characterization shows that most of the thin flakes (Figures 1b, 1c, and 1d) are few-layer
7

materials, such as tri-layers (Figures 2a and 2b) and four-layers (Supporting Information Figure S2), as evident from the cross-section height profiles (bottom parts of Figures 2a and 2b). More importantly, we frequently observed that there are ribbon-like features lying on top of these thin flakes (Supporting Information Figure S3). The ribbons usually have heights equal to one or two-layer thickness of $WSe_2$ (red and black height profiles in Figures 2a and 2b). Figures 2c and 2d are Raman spectra taken from several such thin flakes. Two characteristic peaks were observed in the region from 245 cm$^{-1}$ to 260 cm$^{-1}$, which can be assigned to $E_{2g}^1$ and $A_{1g}$ modes of $WSe_2$, respectively.[43-45] For few-layer $WSe_2$, these two peaks are very close to each other and thickness-dependent shift of Raman peak position is relatively small comparing to $MoS_2$.[43, 46] Therefore, it is difficult to use the peak-to-peak distance to precisely determine the layer numbers. Nevertheless, the existence of $B_{2g}^1$ peaks at 309 cm$^{-1}$ reveals that they are few-layer flakes,[44, 45] which is consistent with AFM measurements shown above. Moreover, there is no peak found at ~420 cm$^{-1}$, which would correspond to $A_{1g}$ mode of $WS_2$, indicating the absence of sulfur doping or the sulfur concentration is negligible in the as-grown $WSe_2$ flakes. Moreover, we performed transmission electron microscope (TEM) and energy dispersive X-ray spectroscopy (EDX) studies. The as-grown $WSe_2$ flakes were first transferred onto TEM grid, using a method reported previously.[47] Figure S4 (Supporting Information) are the TEM images and a typical EDX spectrum, showing no obvious sulfur peak was found. We have acquired a few spectra from different flakes, and the results are very similar. Quantitative analysis shows that if we only count the three elements of W, Se, and S, the atomic ratio of S is below 0.5%, which is the limit of EDX technique. However, we cannot exclude the possibility that there are trace amount of sulfur-doping in as-grown WSe$_2$ flakes. Compared with recent papers on the growth of TMDC alloys,[48-51] the amount of sulfur as well as its temperature were much lower in our study, which lead to the growth of WSe$_2$ flakes with negligible sulfur doping, if any.

We also performed Raman mapping on the same flake shown in Figure 2b. Figures 2e and 2f are Raman intensity mapping images for the $A_{1g}$ and $B_{2g}^1$ modes of WSe$_2$, respectively. As it can be clearly discerned, the areas covered by ribbons are darker than other parts. The same mapping study was performed on another flake with ribbons on top, and the results are consistent with Figures 2e and 2f (Supporting Information Figure S3b).This suggests that the ribbon-covered parts are thicker than other areas since both $A_{1g}$ and $B_{2g}^1$ peak intensities will decrease with increasing the layer numbers of WSe$_2$.[44,45] The intensity of Raman peaks are determined by the intensity of incident light and the amount of materials involved during the scattering process. For thick WSe$_2$ materials, or more generally high reflective index TMDCs, the local electrical field will be much weaker than the incident electrical field, the so-called local field effect. Therefore, thick WSe$_2$ materials will exhibit weak Raman signal. On the other hand, for very thin WSe$_2$ layers, the local field effect is relatively small, and the Raman intensity will be related to the amount of materials involved during Raman scattering process.[52] Therefore, the overall Raman intensity will be jointed determined by the local filed effect and the amount of materials, and there might be a peak at certain height. In our study, we observed that the intensity of Raman peaks decreased in few-layer samples when increasing the layer numbers. This is consistent with a recent study where they showed that bilayer $WSe_2$ exhibit the strongest Raman intensity, and decrease in the order of 3L, 4L, and 5L.[44]

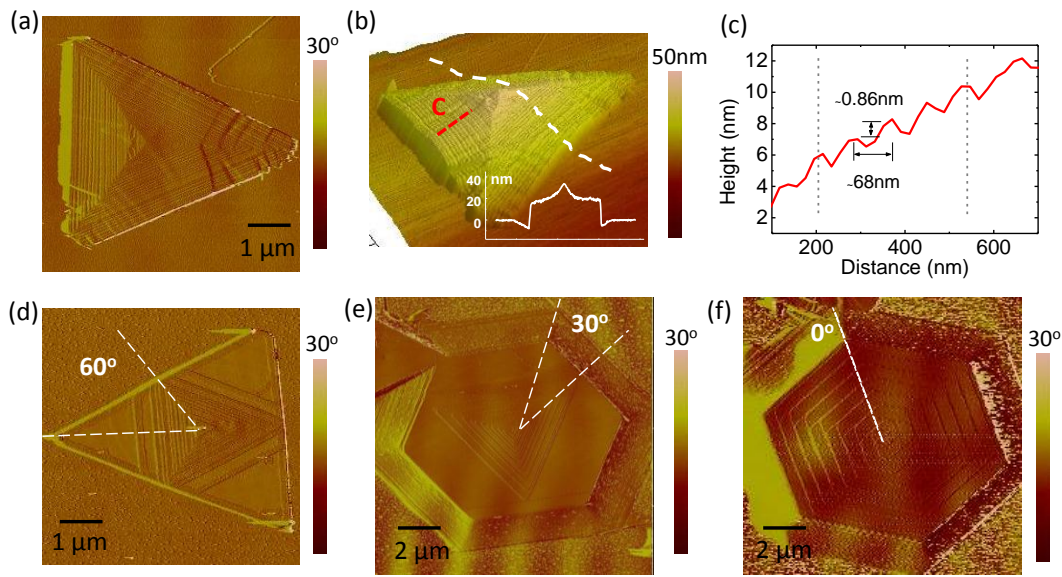

**Figure 3. AFM studies of thick $WSe_2$ flakes with different stacking morphologies. (a) AFM phase image of a thick triangular flake. Helical fringes and herring-bone contours are clearly observed. (b) A 3D AFM image of the flake in (a), showing a pyramid-like structure vividly with a height of ~40 nm from base to the summit. (c) AFM height profile of the pyramid-like flake along the red line in (b). (d, e, f) AFM phase images of three other thick flakes with different morphologies.**

AFM characterization reveals even more interesting features for stacked thick $WSe_2$ flakes. Figures 3a, 3d, 3e, and 3f are AFM phase images of four typical thick flakes with different stacking morphologies. In a more systematical AFM analysis (Supporting Information Figures S5 and S6), we found that flakes with different stacking morphologies appear at different frequencies. Among the ten triangular flakes we examined, six of them have a 0° stacking angle, two have a 60° stacking angle, and one has a 15° stacking angle. We also checked ten hexagonal flakes, eight of them are hexagon-triangle stacks with a 30° stacking angle, and two flakes are hexagon-hexagon stacks. These results of stacking angles and shapes are consistent with what been observed under optical microscopy (Figure 1). Moreover, steps and helical fringes were clearly observed, which strongly support the existence of screw dislocations in these WSe$_2$ samples. Additional evidences like herring-bone contours were also observed. Taking all the optical microscopy, AFM, and Raman observations together (Figures 1, 2, and 3), we proposed that these WSe$_2$ flakes followed a SDD spiral growth fashion.[40, 41] In classical crystal growth theory, there are three basic growth types:[40, 41] SDD growth (BCF theory),[53, 54] layer-by-layer (LBL) growth,[53] and dendritic growth. The growth preference depends on the degree of supersaturation as expressed as σ = ln($c/c_o$), where σ is the degree of supersaturation, $c$ is the precursor concentration, and $c_o$ is the equilibrium concentration.[40, 41, 53] At a low supersaturation (σ) condition, SDD growth is much more favorable than the other two because screw dislocations can provide active edges as nucleation sites, while LBL growth and dendritic growth require nuclei formations that occur only at certain high supersaturation conditions.

To describe the screw dislocations more quantitatively, we measured the key parameters of the screw dislocations in as-grown $WSe_2$ flakes, including step height (h), terrace width (λ), and slope (p = h/λ). These values can reflect the growth conditions at certain degree. In most cases, low supersaturation (σ) would result a small p and large λ.[40] Figure 3b is a three-dimensional (3D) image of the same flake shown in Figure 3a. A pyramid-like structure with a height of 40 nm (from base to summit) is clearly observed. The height profile along the red dash line in Figure 3b is shown in Figure 3c. The step height (h) of this particular dislocation hillock is measured to be ~0.86 nm, which is very close to the thickness of an individual $WSe_2$ layer (composed to Se-W-Se layer and has a height of ~0.7 nm). This value is equal to one elemental Burgers vector.[41] Along with the single helical pattern, we conclude that only a single screw dislocation with one elemental Burgers vector is involved in this particular flake. Other spiral growth modes with different Burgers vectors or multiple screw dislocations also exist. For example, the sample in Figure 3f has two helical patterns, which can be a result of simultaneously growth from two screw dislocations. Another important parameter, terrace width (λ), of the sample in Figures 3a and 3b is ~68 nm, which is smaller comparing to the other flakes in Figures 3d, 3e, and 3f. From screw dislocation growth theory, the terrace width is mostly affected by the reactant concentrations. Specifically, the value of λ is determined by the lateral step velocity ($v_s$) and growth rate normal to the surface ($R_m$). A sample grown with a relative higher $R_m$ and smaller $v_s$ would display a smaller λ. And usually, a smaller λ facilitates a better observation of herring-bone contours than a larger λ does.

This is the reason why Figure 3a has sharp herring-bone contours while the others do not. With the measured terrace width (λ) and step height (h), we can calculate that the flake shown in Figures 3a and 3b has a slope of $p = h/\lambda = 0.0126$. Typically, a small terrace width will lead to a high slope of the pyramid. Moreover, along this study, we have examined tens of such spiral grown flakes and found that both clockwise and counterclockwise grown samples with varied stacking angles exist. The different spiral features are likely to originate from the spiral growth curvature, and the spiral curvature depends much on the reactant concentrations as well. However, such concentration-curvature relation is rather complicated. In other words, the terrace width and curvature do not have direct relations. For example, in our experiments, we find flakes with the same stacking angle, but different terrace widths as shown in Figure S5 and Figure S6. As the concentrations of reactants in a CVD tube furnace varies from location to location due to the use of solid precursors, this leads to the observation of many kinds of stacking morphologies on a same substrate.

To illustrate how a screw dislocation may generate, taking the situation with only one elemental Burgers vector as an example, we drew schematic diagrams to illustrate this process (Figures 4a, 4b, and 4c). When two $WSe_2$ domains intersect, it can cause an uplifting of one grain boundary (Figure 4b). This process will leave some unsaturated Se atoms hanging on this uplifted edge. At this moment, latter sources can be either added to the lateral edges forming a lateral growth or to the uplifted edge forming a second layer growth. Once the second layer is extended, a screw dislocation will be created, which facilitates further spiral growth of WSe$_2$ following a SDD model (Figure 4c). This stage corresponds to what we have observed on those thin flakes in Figures 2a and 2b. Later on, among those three growth modes of LBL growth, continuous lateral growth, and SDD spiral growth, which growth type is preferred depends on the concentration of reactants. As mentioned above, according to classical crystal growth theory, SDD spiral growth is the most favorable type at low supersaturation conditions.[40, 41] For CVD growth of MoS$_2$, MoO$_3$ and S react with each other easily, thus the concentrations of active reactants may be high enough to facilitate a large area lateral growth and few-layer LBL growth. Nevertheless, some similar uplifted second layer features also exist under certain conditions.[42] On the other hand, it is rather different for CVD growth of WSe$_2$. Due to the low reactivity of Se, Huang *et al.* found that H$_2$ has to be introduced to help reducing WO$_3$ into WO$_{3-x}$ and to obtain monolayer dominated WSe$_2$ flakes.[32] Here in our case, we discovered that sulfur can play a similar role as H$_2$ does, and it is not a must to have H$_2$ involved. Since sulfur is not a strong reducer as H$_2$ does, and the amount of sulfur is quite little in our case, the concentration of WO$_{3-x}$ active source is still not high enough for other types of growth except SDD growth in our case. So in this situation, lateral growth and LBL growth are likely to be prohibited while spiral growth on screw dislocations is preferred. We also performed CVD experiments without the addition of sulfur powders, and no such WSe$_2$ growth was found. This is consistent with Huang *et al.*'s recent results.[32]

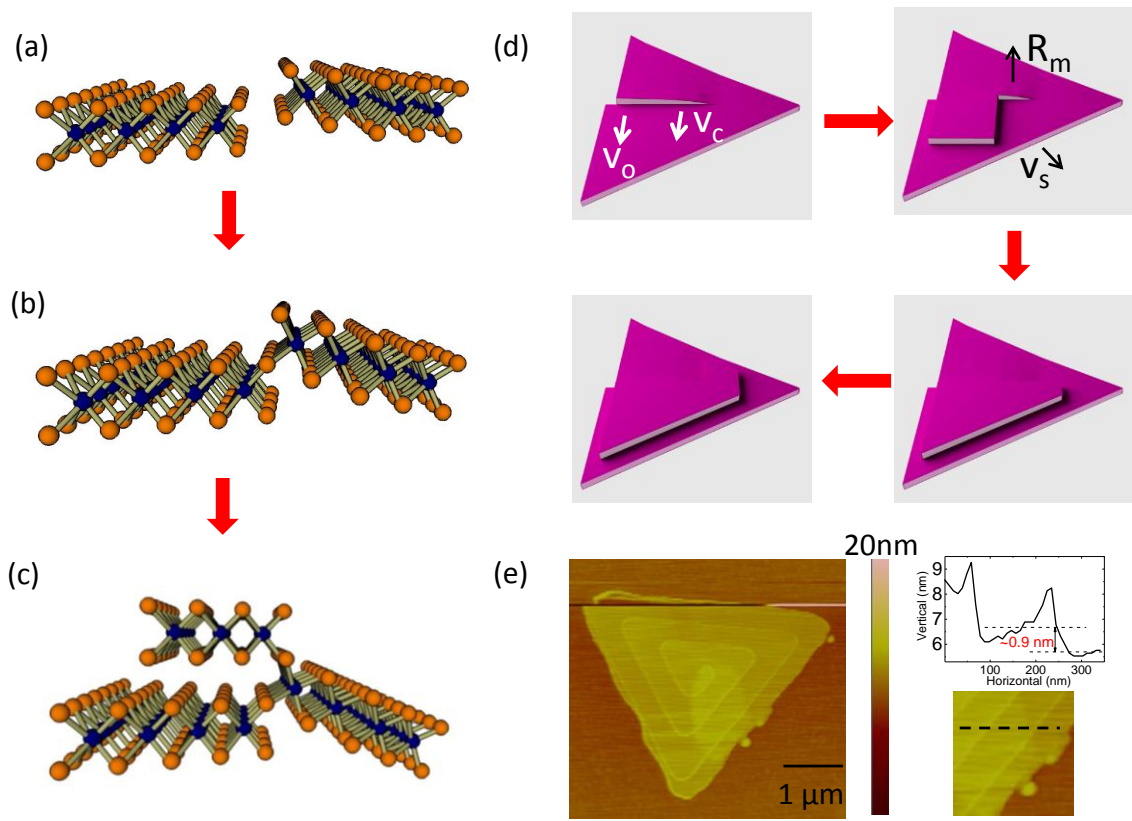

**Figure 4. Proposed models for SDD growth of WSe$_2$. (a) Two adjacent WSe$_2$ domains before intersecting. (b) The boundary uplifting occurs when the two domains intersect. (c) A screw dislocation generated after the second layer extension on the uplifted edge. (d) Schematic diagrams showing the process of screw dislocation propagation. (e) AFM image of a spiral grown flake at early stage. A zoom-in image with the height profile shown in the right part.**

Figure 4d illustrates how a screw dislocation propagates and eventually leads to the pyramid-like WSe$_2$ flake. At first with a low supersaturation environment, the step propagation velocities at the dislocation core ($v_c$) and outer edges ($v_o$) are approximately the same. This is evidenced by the uniformity of terrace width ($\lambda$) in AFM images. Thus, the steps can continuously spread without piling up. Moreover, the growth rate normal to the surface ($R_m$) is much lower than lateral step velocities ($v_s$), which leads to a 2D flake instead of a 1D structure.[40] A product at the early stage with only a few terrace steps was shown in Figure 4e, which clearly reveals the early structures of screw dislocations in the samples. During the CVD growth period, the concentration of Se will decrease gradually along with the Ar flow due to the amount of Se powder left will decrease. Based on SDD growth theory, the terrace width will increase as the reactant concentration decreases. This may be the main reason that we frequently observe on most flakes that the terrace width increases after certain periods.

We further performed electrical transport measurements to study the electronic quality of the as-grown $WSe_2$ flakes. Compared to a recent study of CVD growth of $WSe_2$ on sapphire,[32] our samples were grown directly on $Si/SiO_2$ substrates, which facilitates the fabrication of back-gated field-effect transistors (FETs). The devices were fabricated using standard e-beam lithography and the electrodes were 1 nm/75 nm of Ti/Pd (see Methods for details). Figure 5a shows a device with a 2 μm channel length on a 5-nm-thick flake and Figure 5b shows another device with a 1 μm channel length on a 20-nm-thick flake. The $I_{ds}$-$V_g$ family curves are shown in each corresponding figures (Figures 5c and 5d). Interestingly, since Pd was used as the contacts, these devices show unipolar p-type behavior. This is different with recent results of ambipolar transport behavior of $WSe_2$ flakes when using Au as contact.[14, 15] The results suggest that the transport behavior of WSe$_2$ can be tuned by careful selection of metal contacts, as also demonstrated in MoS$_2$ devices.[55] The on/off current ratios of the devices are around $10^4$ to $10^6$, and effective hole mobility is about 40cm$^2$/V·s. These values compared favorably with recent reported values for few-layer WSe$_2$.[10, 56]

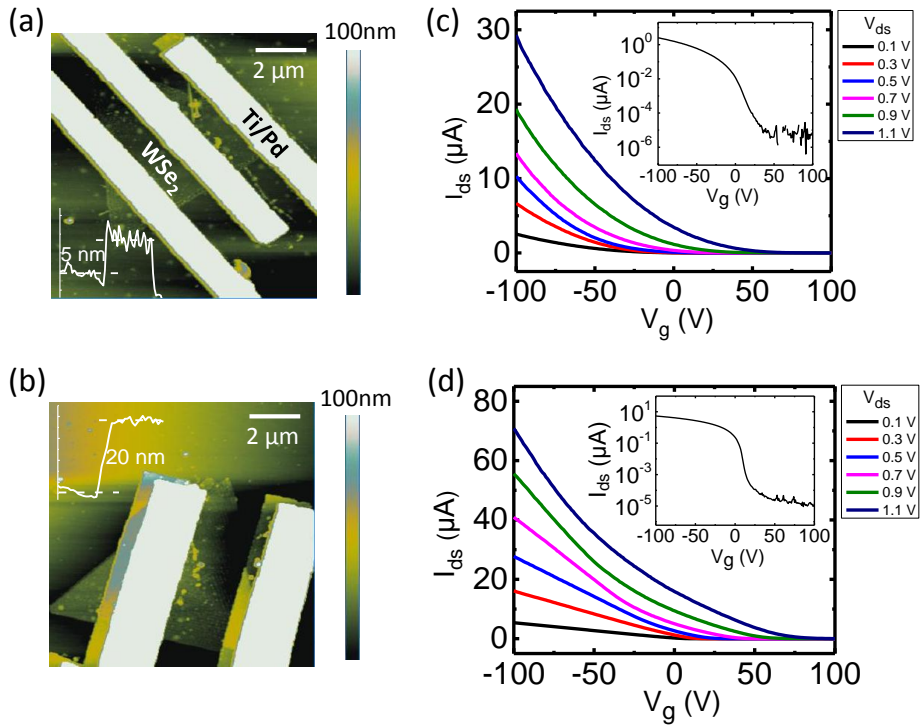

**Figure 5. Device performance of CVD-grown WSe$_2$ flakes.** (a) AFM image of a back-gated WSe$_2$ FET on a 5-nm-thick sample along with its I$_{ds}$-V$_g$ family curves at different V$_{ds}$ (c). (b) Another back-gated WSe$_2$ FET on a 20-nm-thick sample with its I$_{ds}$-V$_g$ family curves (d). Insets in the right plots (c) and (d) are transfer curves plotted in a log scale at V$_{ds}$ = 0.1 V.

**Conclusion:**

In conclusion, few-layer and pyramid-like WSe$_2$ flakes were synthesized using a sulfur-assisted CVD method. The WSe$_2$ growth was proposed to follow a SDD growth process due to the low supersaturation of reactants. Sulfur was found to play an important role here as it can partially reduce the WO$_3$ while keeping the WO$_{3-x}$ concentration low. Key screw dislocation features including steps, helical fringes, and herring-bone contours were observed to support the spiral growth. Schematic models were drawn to illustrate how the screw dislocations generate and propagate. In addition, transistors fabricated using these WSe$_2$ flakes show that they possess decent device performance with on/off current ratio up to $10^6$ and hole mobility up to 44 cm$^2$/V·s. This work sheds new light on the understanding of growth mechanism of layered WSe$_2$, which might be also adoptable in other 2D materials.

**Methods:**

**CVD growth of WSe$_2$ flakes.** In a typical experiment, WO$_3$ nano-powders (40 mg, Sigma Aldrich) were loaded in a quartz boat and put at the center of a 1-inch tube furnace. A Si/SiO$_2$ substrate (300 nm SiO$_2$) was placed on top of the quartz boat at a distance ~0.5 cm at the downstream of the WO$_3$ powders. Selenium powders (30 mg, Sigma Aldrich) and a small amount of sulfur powders (5 mg, Sigma Aldrich) were put at upstream with temperatures of 400 °C (for selenium) and 85 °C (for sulfur). Since the temperature was below the melting point of sulfur (115.2 °C), only very little sulfur vapor was carried to the reaction zone by a small flow of Ar gas (14 sccm). This is evident from the fact that there is still some sulfur left in the boat after the experiments. The furnace was first purged with 100 sccm of Ar for 10 minutes, then raised to the growth temperatures in 16 minutes, and kept for 12 minutes for $WSe_2$ growth. After that, it was naturally cooled down to room temperature. The flow rate of Ar was 14 sccm during the CVD process.

**Characterization.** The as-grown flakes were characterized by optical microscopy, Raman spectroscopy (532 nm laser, Renishaw Raman), AFM (DI 3100 Digital Instruments, tapping mode), TEM (JEOL 2100 F, 200 kV, equipped with detector for EDX).

**Device Fabrication and Measurements.** The transistors were directly fabricated on $Si/SiO_2$ substrates where $WSe_2$ were grown on using standard e-beam lithography. A PMMA layer was first spin-coated onto the $Si/SiO_2$ surface. Then, e-beam lithography was conducted to pattern source/drain electrodes, followed by develop, metal deposition and lift-off processes. Ti/Pd electrodes (1 nm/75 nm) were deposited at $1 \times 10^{-6}$ Torr using an e-beam evaporator. The device measurements were performed under Aglient 4516 in ambient condition.

*Conflict of Interest:* The authors declare no competing financial interest.

**Acknowledgements:**

This work was supported by the Office of Naval Research (ONR) and the Air Force Office of Scientific Research (AFOSR). We would like to acknowledge the collaboration of this research with King Abdul-Aziz City for Science and Technology (KACST) *via* The Center of Excellence for Nanotechnologies (CEGN). TEM and EDX images used in this article were acquired at the Center for Electron Microscopy and Microanalysis, University of Southern California. We acknowledge Professor Stephen Cronin of University of Southern California for access to Raman facility.**Supporting Information:**

Additional AFM, TEM, EDX, and Raman results. This material is available free of charge *via* the Internet at http://pubs.acs.org.

**References**

1. Wang, Q. H.; Kalantar-Zadeh, K.; Kis, A.; Coleman, J. N.; Strano, M. S. Electronics and Optoelectronics of Two-dimensional Transition Metal Dichalcogenides. *Nat. Nanotechnol.* **2012**, *7*, 699-712.
2. Chhowalla, M.; Shin, H. S.; Eda, G.; Li, L. J.; Loh, K. P.; Zhang, H. The Chemistry of Two-dimensional Layered Transition Metal Dichalcogenide Nanosheets. *Nat. Chem.* **2013**, *5*, 263-275.
3. Geim, A. K.; Grigorieva, I. V. Van der Waals Heterostructures. *Nature* **2013**, *499*, 419-425.
4. Radisavljevic, B.; Radenovic, A.; Brivio, J.; Giacometti, V.; Kis, A. Single-layer $MoS_2$ Transistors. *Nat. Nanotechnol.* **2011**, *6*, 147-150.
5. Kuc, A.; Zibouche, N.; Heine, T. Influence of Quantum Confinement on the Electronic Structure of the Transition Metal Sulfide $TS_2$. *Phys. Rev. B* **2011**, *83*, 245213.
6. Mak, K. F.; He, K.; Shan, J.; Heinz, T. F. Control of Valley Polarization in Monolayer $MoS_2$ by Optical Helicity. *Nat. Nanotechnol.* **2012**, *7*, 494-498.
7. Zeng, H.; Dai, J.; Yao, W.; Xiao, D.; Cui, X. Valley Polarization in $MoS_2$ Monolayers by Optical Pumping. *Nat. Nanotechnol.* **2012**, *7*, 490-493.
8. Pu, J.; Yomogida, Y.; Liu, K. K.; Li, L. J.; Iwasa, Y.; Takenobu, T. Highly Flexible $MoS_2$ Thin-film Transistors with Ion Gel Dielectrics. *Nano Lett.* **2012**, *12*, 4013-4017.
9. Chang, H.-Y.; Yang, S.; Lee, J.; Tao, L.; Hwang, W.-S.; Jena, D.; Lu, N.; Akinwande, D. High-Performance, Highly Bendable $MoS_2$ Transistors with High-K Dielectrics for Flexible Low-Power Systems. *ACS Nano* **2013**, *7*, 5446-5452.

10. Tosun, M.; Chuang, S.; Fang, H.; Sachid, A. B.; Hettick, M.; Lin, Y.; Zeng, Y.; Javey, A. High-Gain Inverters Based on WSe$_2$ Complementary Field-Effect Transistors. *ACS Nano* **2014**, *8*, 4948-4953.

11. Wang, H.; Yu, L.; Lee, Y. H.; Shi, Y.; Hsu, A.; Chin, M. L.; Li, L. J.; Dubey, M.; Kong, J.; Palacios, T. Integrated Circuits Based on Bilayer MoS$_2$ Transistors. *Nano Lett.* **2012**, *12*, 4674-4680.

12. Lopez-Sanchez, O.; Lembke, D.; Kayci, M.; Radenovic, A.; Kis, A. Ultrasensitive Photodetectors Based on Monolayer MoS$_2$. *Nat. Nanotechnol.* **2013**, *8*, 497-501.

13. Ross, J. S.; Klement, P.; Jones, A. M.; Ghimire, N. J.; Yan, J.; Mandrus, D. G.; Taniguchi, T.; Watanabe, K.; Kitamura, K.; Yao, W.; *et al.* Electrically Tunable Excitonic Light-emitting Diodes Based on Monolayer WSe$_2$ p-n Junctions. *Nat. Nanotechnol.* **2014**, *9*, 268-272.

14. Pospischil, A.; Furchi, M. M.; Mueller, T. Solar-energy Conversion and Light Emission in an Atomic Monolayer p-n Diode. *Nat. Nanotechnol.* **2014**, *9*, 257-261.

15. Baugher, B. W. H.; Churchill, H. O. H.; Yang, Y.; Jarillo-Herrero, P. Optoelectronic Devices Based on Electrically Tunable p-n Diodes in a Monolayer Dichalcogenide. *Nat. Nanotechnol.* **2014**, *9*, 262-267.

16. Chang, K.; Chen, W. L-Cysteine-Assisted Synthesis of Layered MoS$_2$/Graphene Composites with Excellent Electrochemical Performances for Lithium Ion Batteries. *ACS Nano* **2011**, *5*, 4720-4728.

17. Hinnemann, B.; Moses, P. G.; Bonde, J.; Jørgensen, K. P.; Nielsen, J. H.; Horch, S.; Chorkendorff, I.; Nørskov, J. K. Biomimetic Hydrogen Evolution: MoS$_2$ Nanoparticles as Catalyst for Hydrogen Evolution. *J. Am. Chem. Soc.* **2005**, *127*, 5308-5309.

18. Lee, Y. H.; Zhang, X. Q.; Zhang, W.; Chang, M. T.; Lin, C. T.; Chang, K. D.; Yu, Y. C.; Wang, J. T.; Chang, C. S.; Li, L. J.; *et al.* Synthesis of Large-area MoS$_2$ Atomic Layers with Chemical Vapor Deposition. *Adv. Mater.* **2012**, *24*, 2320-2325.

19. Ling, X.; Lee, Y. H.; Lin, Y.; Fang, W.; Yu, L.; Dresselhaus, M. S.; Kong, J. Role of the Seeding Promoter in MoS$_2$ Growth by Chemical Vapor Deposition. *Nano Lett.* **2014**, *14*, 464-472.

20. Wang, X.; Gong, Y.; Shi, G.; Chow, W. L.; Keyshar, K.; Ye, G.; Vajtai, R.; Lou, J.; Liu, Z.; Ringe, E.; *et al*. Chemical Vapor Deposition Growth of Crystalline Monolayer MoSe$_2$. *ACS Nano* **2014**, *8*, 5125–5131.

21. Zhang, Y.; Zhang, Y.; Ji, Q.; Ju, J.; Yuan, H.; Shi, J.; Gao, T.; Ma, D.; Liu, M.; Chen, Y.; *et al*. Controlled Growth of High-Quality Monolayer WS$_2$ Layers on Sapphire and Imaging Its Grain Boundary. *ACS Nano* **2013**, *7*, 8963-8971.

22. Coleman, J. N.; Lotya, M.; O'Neill, A.; Bergin, S. D.; King, P. J.; Khan, U.; Young, K.; Gaucher, A.; De, S.; Smith, R. J.; *et al*. Two-dimensional Nanosheets Produced by Liquid Exfoliation of Layered Materials. *Science* **2011**, *331*, 568-571.

23. Kong, D.; Wang, H.; Cha, J. J.; Pasta, M.; Koski, K. J.; Yao, J.; Cui, Y. Synthesis of MoS$_2$ and MoSe$_2$ Films with Vertically Aligned Layers. *Nano Lett.* **2013**, *13*, 1341-1347.

24. Najmaei, S.; Liu, Z.; Zhou, W.; Zou, X.; Shi, G.; Lei, S.; Yakobson, B. I.; Idrobo, J. C.; Ajayan, P. M.; Lou, J. Vapour Phase Growth and Grain Boundary Structure of Molybdenum Disulphide Atomic Layers. *Nat. Mater.* **2013**, *12*, 754-759.

25. van der Zande, A. M.; Huang, P. Y.; Chenet, D. A.; Berkelbach, T. C.; You, Y.; Lee, G. H.; Heinz, T. F.; Reichman, D. R.; Muller, D. A.; Hone, J. C. Grains and Grain Boundaries in Highly Crystalline Monolayer Molybdenum Disulphide. *Nat. Mater.* **2013**, *12*, 554-561.

26. Lauritsen, J. V.; Kibsgaard, J.; Helveg, S.; Topsoe, H.; Clausen, B. S.; Laegsgaard, E.; Besenbacher, F. Size-dependent Structure of MoS$_2$ Nanocrystals. *Nat. Nanotechnol.* **2007**, *2*, 53-58.

27. Zhu, W.; Low, T.; Lee, Y. H.; Wang, H.; Farmer, D. B.; Kong, J.; Xia, F.; Avouris, P. Electronic Transport and Device Prospects of Monolayer Molybdenum Disulphide Grown by Chemical Vapour Deposition. *Nat. Commun.* **2014**, *5*, 3087.

28. Duerloo, K. A.; Li, Y.; Reed, E. J. Structural Phase Transitions in Two-dimensional Mo- and W-dichalcogenide Monolayers. *Nat. Commun.* **2014**, *5*, 4214.

29. Perkins, F. K.; Friedman, A. L.; Cobas, E.; Campbell, P. M.; Jernigan, G. G.; Jonker, B. T. Chemical Vapor Sensing with Monolayer $MoS_2$. *Nano Lett.* **2013**, *13*, 668-673.

30. Liu, B.; Chen, L.; Liu, G.; Abbas, A. N.; Fathi, M.; Zhou, C. High-Performance Chemical Sensing Using Schottky-Contacted Chemical Vapor Deposition Grown Monolayer $MoS_2$ Transistors. *ACS Nano* **2014**, *8*, 5304-5314.

31. Balendhran, S.; Ou, J. Z.; Bhaskaran, M.; Sriram, S.; Ippolito, S.; Vasic, Z.; Kats, E.; Bhargava, S.; Zhuiykovd, S.; Kalantar-zadeh, K. Atomically Thin Layers of $MoS_2$ *via* a Two Step Thermal Evaporation–Exfoliation Method. *Nanoscale* **2012**, *4*, 461-466.

32. Huang, J.-K.; Pu, J.; Hsu, C.-L.; Chiu, M.-H.; Juang, Z.-Y.; Chang, Y.-H.; Chang, W.-H.; Iwasa, Y.; Takenobu, T.; Li, L.-J. Large-Area Synthesis of Highly Crystalline $WSe_2$ Monolayers and Device Applications. *ACS Nano* **2014**, *8*, 923-930.

33. Xu, K.; Wang, Z.; Du, X.; Safdar, M.; Jiang, C.; He, J. Atomic-layer Triangular $WSe_2$ Sheets: Synthesis and Layer-dependent Photoluminescence Property. *Nanotechnology* **2013**, *24*, 465705.

34. Lin, Y.-C.; Lu, N.; Perea-Lopez, N.; Lin, J. L. Z.; Peng, X.; Lee, C. H.; Sun, C.; Calderin, L.; Browning, P. N.; Bresnehan, M. S.; *et al.* Direct Synthesis of van der Waals Solids. *ACS Nano* **2014**, *8*, 3715-3723.

35. Meng, F.; Jin, S. The Solution Growth of Copper Nanowires and Nanotubes Is Driven by Screw Dislocations. *Nano Lett.* **2012**, *12*, 234-239.

36. Morin, S. A.; Bierman, M. J.; Tong, J.; Jin, S. Mechanism and Kinetics of Spontaneous Nanotube Growth Driven by Screw Dislocations. *Science* **2010**, *328*, 476-480.

37. Morin, S. A.; Jin, S. Screw Dislocation-driven Epitaxial Solution Growth of ZnO Nanowires Seeded by Dislocations in GaN Substrates. *Nano Lett.* **2010**, *10*, 3459-3463.

38. Zhu, J.; Peng, H.; Marshall, A. F.; Barnett, D. M.; Nix, W. D.; Cui, Y. Formation of Chiral Branched Nanowires by the Eshelby Twist. *Nat. Nanotechnol.* **2008**, *3*, 477-481.

39. Bierman, M. J.; Lau, Y. K.; Kvit, A. V.; Schmitt, A. L.; Jin, S. Dislocation-driven Nanowire Growth and Eshelby Twist. *Science* **2008**, *320*, 1060-1063.

40. Morin, S. A.; Forticaux, A.; Bierman, M. J.; Jin, S. Screw Dislocation-driven Growth of Two-dimensional Nanoplates. *Nano Lett.* **2011**, *11*, 4449-4455.

41. Zhuang, A.; Li, J. J.; Wang, Y. C.; Wen, X.; Lin, Y.; Xiang, B.; Wang, X.; Zeng, J. Screw-dislocation-driven Bidirectional Spiral Growth of $Bi_2Se_3$ Nanoplates. *Angew. Chem.* **2014**, *53*, 6425-6429.

42. Ji, Q.; Zhang, Y.; Gao, T.; Zhang, Y.; Ma, D.; Liu, M.; Chen, Y.; Qiao, X.; Tan, P. H.; Kan, M.; *et al.* Epitaxial Monolayer $MoS_2$ on Mica with Novel Photoluminescence. *Nano Lett.* **2013**, *13*, 3870-3877.

43. Luo, X.; Zhao, Y.; Zhang, J.; Toh, M.; Kloc, C.; Xiong, Q.; Quek, S. Y. Effects of Lower Symmetry and Dimensionality on Raman Spectra in Two-dimensional $WSe_2$. *Phys. Rev. B* **2013**, *88*, 195313.

44. Li, H.; Lu, G.; Wang, Y.; Yin, Z.; Cong, C.; He, Q.; Wang, L.; Ding, F.; Yu, T.; Zhang, H. Mechanical

Exfoliation and Characterization of Single- and Few-layer Nanosheets of WSe$_2$, TaS$_2$, and TaSe$_2$. *Small* **2013**, *9*, 1974-81.

45. Tonndorf, P.; Schmidt, R.; Bottger, P.; Zhang, X.; Borner, J.; Liebig, A.; Albrecht, M.; Kloc, C.; Gordan, O.; Zahn, D. R.; *et al*. Photoluminescence Emission and Raman Response of Monolayer MoS$_2$, MoSe$_2$, and WSe$_2$. *Opt. Express* **2013**, *21*, 4908-4916.

46. Sahin, H.; Tongay, S.; Horzum, S.; Fan, W.; Zhou, J.; Li, J.; Wu, J.; Peeters, F. Anomalous Raman Spectra and Thickness-dependent Electronic Properties of WSe$_2$. *Phys. Rev. B* **2013**, *87*, 165409.

47. Liu, B.; Ren, W.; Gao, L.; Li, S.; Pei, S.; Liu, C.; Jiang, C.; Cheng, H.M. Metal-Catalyst-Free Growth of Single-Walled Carbon Nanotubes. *J. Am. Chem. Soc.* **2009**, *131*, 2082-2083.

48. Li, H.; Duan, X.; Wu, X.; Zhuang, X.; Zhou, H.; Zhang, Q.; Zhu, X.; Hu, W.; Ren, P.; Guo, P.; *et al.* Growth of Alloy MoS$_{2x}$Se2$_{(1-x)}$ Nanosheets with Fully Tunable Chemical Compositions and Optical Properties. *J. Am. Chem. Soc.* **2014**, *136*, 3756-3759.

49. Mann, J.; Ma, Q.; Odenthal, P. M.; Isarraraz, M.; Le, D.; Preciado, E.; Barroso, D.; Yamaguchi, K.; von Son Palacio, G.; Nguyen, A.; *et al.* 2-dimensional Transition Metal Dichalcogenides with Tunable Direct Band Gaps: MoS$_{2(1-x)}$Se$_{2x}$ Monolayers. *Adv. Mater.* **2014**, *26*, 1399-1404.

50. Gong, Y.; Liu, Z.; Lupini, A. R.; Shi, G.; Lin, J.; Najmaei, S.; Lin, Z.; Elias, A. L.; Berkdemir, A.; You, G.; *et al.* Band Gap Engineering and Layer-by-Layer Mapping of Selenium-Doped Molybdenum Disulfide. *Nano lett.* **2014**, *14*, 442-449.

51. Chen, Y.; Xi. J.; Dumcenco, D. O.; Liu, Z.; Suenaga, K.; Wang, D.; Shuai, Z.; Huang, Y.; Xie, L. Tunable Band Gap Photoluminescence from Atomically Thin Transition-Metal Dichalcogenide Alloys. *ACS Nano* **2013**, *7*, 4610-4616.

52. Splendiani, A.; Sun, L.; Zhang, Y.; Li, T.; Kim, J.; Chim, C. Y.; Galli, G.; Wang, F. Emerging Photoluminescence in Monolayer MoS$_2$. *Nano lett.* **2010**, *10*, 1271-1275.

53. Jin, S.; Bierman, M. J.; Morin, S. A. A New Twist on Nanowire Formation: Screw-Dislocation-Driven Growth of Nanowires and Nanotubes. *J. Phys. Chem. Lett.* **2010**, *1*, 1472-1480.

54. Burton, W. K.; Cabrera, N.; Frank, F. C. The Growth of Crystals and the Equilibrium Structure of their Surfaces. *Philosophical Transactions of the Royal Society A: Mathematical, Physical and Engineering Sciences* **1951**, *243*, 299-358.

55. Chuang, S.; Battaglia, C.; Azcatl, A.; McDonnell, S.; Kang, J. S.; Yin, X.; Tosun, M.; Kapadia, R.; Fang, H.; Wallace, R. M.; *et al*. MoS$_2$ P-type transistors and diodes enabled by high work function MoO$_x$ contacts. *Nano Lett*. **2014**, *14*, 1337-1342.

56. Allain, A.; Kis, A. Electron and Hole Mobilities in Single-Layer WSe$_2$. *ACS Nano* **2014**, *8*, 7180–7185.

TOC

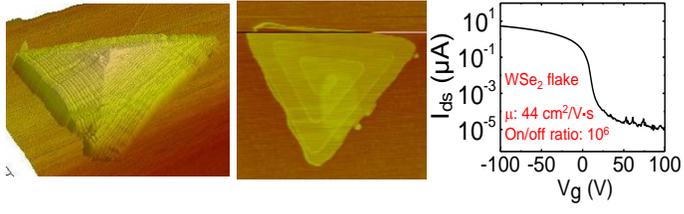